\documentclass[12pt]{article}
\begin{document}
\baselineskip 6mm
\thispagestyle{empty}

\begin{center}
    {\Large  {\bf  Braid Structure and Raising-Lowering Operator }}
\vspace*{3mm} \\
    {\Large {\bf  Formalism in Sutherland Model}}
\vspace*{7mm} \\ 
     {\large Akira Takamura\footnote{
         e-mail address: takamura@eken.phys.nagoya-u.ac.jp
                              }
      and~~Ken'ichi Takano}${}^{\dagger}$
\vspace*{5mm} \\
     ${}^*$
         \sl{Department of Physics, Nagoya University,} \\
         \it{Nagoya 464-8602, Japan}
\vspace*{5mm} \\
     ${}^{\dagger}$
       \sl{Laboratory of Theoretical Condensed Matter Physics and \\
       Research Center for Advanced Photon Technology, \\
       Toyota Technological Institute,} \\
       \it{ Nagoya 468-8511, Japan}
\end{center}
\vspace*{3mm}

\date{}

\begin{abstract}
      We algebraically construct the Fock space of the Sutherland 
model in terms of the eigenstates of the pseudomomenta as 
basis vectors. 
      For this purpose, we derive the raising and lowering operators 
which increase and decrease eigenvalues of pseudomomenta. 
      The operators exchanging eigenvalues of two pseudomomenta 
have been known. 
      All the eigenstates are systematically produced by starting 
from the ground state and multiplying these operators to it. 
\end{abstract}

\hspace*{\parindent}


      The Sutherland model is a solvable quantum many-body system 
with inverse-square interaction on a circumference \cite{Sutherland}. 
     The ground-state wave function is of the Jastrow type and excited 
states are polynomials multiplied by the ground state. 
      Among the polynomials, the symmetric ones are Jack 
polynomials \cite{Jack,Stanley,Macdonald}, while 
the others are called nonsymmetric Jack polynomials. 
     These energy eigenstates can be taken as eigenstates of the 
pseudomomenta \cite{Dunkl,Cherednik}, which commute 
with each other and with the Hamiltonian. 

      For its rich content, the Sutherland model has been zealously 
investigated at various standpoints. 
      For example, the Sutherland model is regarded as a model 
which describes the edge state in the fractional quantum Hall 
effect \cite{Kawakami}. 
     It may  describe the fractional statistics of quasiparticles 
\cite{Laughlin}. 
      Also a deep connection of this model to the conformal field 
theory is found \cite{Yamada}. 
      Haldane argued that the Sutherland model is equivalent to 
the system of particles obeying the exclusion statistics 
if the coupling constant is a rational number \cite{Haldane1}. 
      Based on this assumption he obtained the concrete form of 
the two point correlation function; i.e., as intermediate states, 
he only used free particle states obeying the exclusion statistics. 
      The result coincides with the exact one which was calculated 
by using the duality of Jack polynomials 
\cite{Haldane2,Simons,Haldane3,Ha}.  
      The duality means the invariance of the Jack polynomials 
under a nonlinear transformation with the replacement of 
the coupling constant by its inverse. 
      In the Sutherland model, many interesting properties 
such as the exclusion statistics are deduced by directly inspecting 
the Jack polynomials. 

      To deeply understand the Sutherland model, we need to 
reformulate algebraically the eigenvalue problem of this model.  
      We mention its importance by recalling the case of 
a harmonic oscillator. 
      Although this problem is solved in terms of Hermite polynomials, 
the algebraic approach using creation and annihilation operators 
revealed the essence of the model. 
      The quantum field theory is formulated on the basis of 
harmonic oscillators. 
      In the Calogero model, with inverse-square interaction and 
harmonic potential, creation and annihilation operators 
are examined \cite{Ujino,Kakei}. 
      In the Sutherland model, a hopeful algebraic approach means 
that a simple and transparent algebra determines all the energy 
levels and their degeneracy. 
      There are some algebraic treatments for symmetric \cite{Lapointe} 
and nonsymmetric Jack polinomials \cite{Knop,Kirillov}, 
where a polynomial generates another one by some operations. 
      However such a generated state is not an eigenstate of 
the pseudomomenta except for special cases and is not simple 
for the present purpose to seek a physical transparency. 

      In this letter, we propose a novel algebraic formalism for 
the eigenvalue problem in the Sutherland model. 
      The formalism is based on operators which increase, decrease 
and exchange the eigenvalues of psudemomenta. 
      The raising and lowering operators are derived in this letter 
and the operator for exchange has been introduced \cite{Kirillov}. 
      Starting from the ground state, we can reach an arbitrary 
eigenstate of the pseudomomenta by multiplying a finite number 
of operators.  
      The Fock space of the Sutherland model is reproduced 
in terms of eigenstates of the pseudomomenta.


      We consider $N$ particles on a circumference with length $\pi$ 
and denote the coordinate of the $i$th particle by $\theta_i$. 
      For these particles we introduce an operator $K_{ij}$ ($i \ne j$) 
which exchanges coordinates $\theta_i$ and $\theta_j$; i.e. 
$K_{ij} \theta_i = \theta_j K_{ij}$. 
      Then an extended version of the Sutherland model is given 
by the Hamiltonian 
\begin{equation}
      H = - \sum_{i=1}^{N}\frac{\partial^2}{\partial \theta_i^2}
            + \frac{1}{2}\sum_{i \ne j}
      \frac{\beta(\beta-K_{ij})}{\sin^2[(\theta_i-\theta_j)/2]} , 
\label{Hamiltonian-1}
\end{equation}
where $\beta$ is the coupling constant. 
     This Hamiltonian is invariant against the exchange of the 
coordinates of particles and satisfies the commutation relation 
$[H, K_{ij}] = 0$. 
To make the description simple, we use the complex 
coordinate $z_{i} = \exp(i\theta_i)$ instead of $\theta_i$. 
      The momentum is accordingly represented as  
\begin{equation}
      p_i = z_i \frac{\partial}{\partial z_i}. 
\end{equation}
The quantization condition is then given by 
\begin{equation}
      [p_i, z_j] = \delta_{ij} z_i . 
\label{QC}
\end{equation}
      The Hamiltonian (\ref{Hamiltonian-1}) is rewritten as 
\begin{equation}
      H = \sum_{i=1}^N p_i^2 + \sum_{i,j}
      \frac{z_i z_j}{(z_i -z_j)^2}\beta(\beta-K_{ij}) 
\label{Hamiltonian-2}.
\end{equation}

      Dunkl \cite{Dunkl} and Cherednik \cite{Cherednik} introduced 
the pseudomomentum which is defined as 
\begin{equation}
      D_i = p_i + \beta \sum_{j(>i)} \frac{z_i}{z_i - z_j}K_{ij} 
            - \beta \sum_{j(<i)} K_{ij}\frac{z_i}{z_i - z_j} .
\end{equation}
      In terms of \{$D_i$\}, the Hamiltonian and the total momentum 
are written as
\begin{equation}
      H = \sum_{i=1}^N D_i^2, \quad P = \sum_{i=1}^N D_i 
\label{Hamiltonian-3}.
\end{equation}
      The pseudomomenta are hermitian ($D_i^{\dagger} = D_i$) and 
commute with each other: 
\begin{equation}
      [D_i, D_j] = 0 . \label{Hecke-1}
\end{equation}
      Hence they also commute with the Hamiltonian ($[H, D_i] = 0$). 
      The exchange operators affect the pseudomomenta through 
the relations 
\begin{eqnarray}
   & & D_i K_{i,i+1} - K_{i,i+1} D_{i+1} = \beta , \label{Hecke-2} \\
   & & {}[ D_j, K_{i,i+1}] = 0 . \quad (j \ne i,i+1) \label{Hecke-3}
\end{eqnarray}
      The quantization condition (\ref{QC}) is represented as  
\begin{equation}
       [D_i,z_j] = \left\{
   \begin{array}{rl}
       z_i + \beta z_i \displaystyle{\sum_{j(<i)}} K_{ij}
       + \beta \displaystyle{\sum_{j(>i)}} K_{ij} z_i & \quad (i=j) \\
       - \beta \{
            z_j K_{ij}\theta(i-j) + K_{ij}z_j \theta(j-i)
          \} .  & \quad (i \ne j)
   \end{array} \right.
\label{CR}
\end{equation}
      Here the step function $\theta(x)$ is 1 for $x \ge 0$ and 
is 0 otherwise.
      While the Hamiltonian (\ref{Hamiltonian-3}) is of the form 
for free particles with momenta \{$D_i$\}, the quantization 
condition (\ref{CR}) is rather complicated. 
      That is, all the effects of the long-range interaction 
are involved in the quantization condition (\ref{CR}). 
      For this reason the interaction in the Hamiltonian 
(\ref{Hamiltonian-1}) is called a statistical interaction. 
      The operators \{$D_i,z_j,K_{kl}$\} are closed with respect to 
their mutual products, and thereby forming an algebra. 
      However it is not a Lie algebra, since the commutator of some 
operators is no longer represented by a linear combination of the 
operators. 
      Relations (\ref{Hecke-1}) to (\ref{CR}) form a degenerate
 double affine Hecke algebra. 
 The same structure for the Calogero model is examined by 
 Ujino and Wadati \cite{Ujino} and by Kakei \cite{Kakei} . 


      We construct the energy eigenvalues and the eigenstates of 
the Sutherland model in a completely algebraic manner. 
      First of all, we examine the operator $X_{i,i+1}$ defined by 
\begin{eqnarray}
      X_{i,i+1} &=& i[D_i, K_{i,i+1}] , \quad (i=1,\cdots,N-1) 
\label{definitionX}
\end{eqnarray}
which is clearly hermitian ($X_{i,i+1}^{\dagger} = X_{i,i+1}$). 
      We call this the {\it braid-exclusion operator}. 
      The $q$-deformed version of this operator was first introduced 
by Killirov and Noumi \cite{Kirillov}. 
      The relations (\ref{Hecke-2}) and (\ref{Hecke-3}) for
$D_i$ and $K_{i,i+1}$  are converted to the following relations: 
\begin{eqnarray}
      D_i X_{i,i+1} &=& X_{i,i+1} D_{i+1} , \label{GEO-1} \\
      D_{i+1} X_{i,i+1} &=& X_{i,i+1} D_i , \label{GEO-2} \\
      {}[ D_k, X_{i,i+1}] &=& 0 . \quad (k \ne i,i+1) \label{GEO-3}
\end{eqnarray}
      These equations mean that $X_{i,i+1}$ exchanges the 
pseudomomenta $D_i$ and $D_{i+1}$. 

      From the definition (\ref{definitionX}) the square of $X_{i,i+1}$ 
is written as 
\begin{equation}
      X_{i,i+1}^2 = (D_i-D_{i+1})^2-\beta^2 . 
\label{Exclusion}
\end{equation}
     The positive semidefiniteness of $X_{i,i+1}^2$ requires that 
the difference of eigenvalues of $D_i$ and $D_{i+1}$ must differ 
by a number lager than or equal to $|\beta|$. 
      As will be clear by later examination, any eigenvalues of 
the pseudomomenta are integers in both the special cases of 
$|\beta|$ = 0 and 1. 
      For $|\beta|$ = 0, the particles are bosonic since 
(\ref{Exclusion}) shows that their eigenvalues can take the same value. 
      On the other hand, for $|\beta|$ = 1, the particles are fermionic 
since the eigenvalues must take different integers due to 
(\ref{Exclusion}). 
      Thus the relation (\ref{Exclusion}) for $0 < |\beta| < 1$ shows 
neither bosonic nor fermionic statistics but suggests 
Haldane's exclusion statistics \cite{Haldane4,Haldane5}. 

       The braid-exclusion operators satisfy 
the following relations: 
\begin{eqnarray}
X_{i,i+1}X_{i+1,i+2}X_{i,i+1}
 &=& X_{i+1,i+2}X_{i,i+1}X_{i+1,i+2} \label{Braid-1} , \\
X_{i,i+1}X_{j,j+1} &=& X_{j,j+1}X_{i,i+1}, \quad (|i-j| \geq 2)
\label{Braid-2}
\end{eqnarray}
which are derived from the definition (\ref{definitionX}) and 
the relation (\ref{Hecke-1}) to (\ref{Hecke-3}) by straightforward 
calculation.
      Equations (\ref{Braid-1}) and (\ref{Braid-2}) are the very
 relations which generators of a braid group satisfy \cite{Wilczek}; 
      equation (\ref{Braid-1}) is also of the same form as the 
Yang-Baxter relation. 
      They essentially determine the characters of operators which 
will be introduced below.
      Thus the exchange operator $X_{i,i+1}$ for the pseudomomenta 
possesses both the characters of the exclusion statistics and the 
braid group structure. 
      This is the reason why we have called them braid-exclusion 
operators. 
      The operator however has no inverse operator against any true 
generators for a braid group. 
      In fact the exclusion character (\ref{Exclusion}) allows that 
the eigenvalue of $X_{i,i+1}$ vanishes when the eigenvalue of 
$D_i$ differs from that of $D_{i+1}$ by $\pm \beta$. 


       Next we recall an operator $e^{\dagger}$ which is defined as 
\begin{equation}
      e^{\dagger} = K_{N,N-1}K_{N-1,N-2} \cdots K_{32}K_{21}z_1 ,
\end{equation}
and call it the {\it displacement operator}.
      It was introduced by Knop and Sahi \cite{Knop} 
in relation to nonsymmetric Jack polynomials.  
      Equation $|z_i| = 1$ guarantees its unitarity: 
\begin{equation}
      e^{\dagger}e = e e^{\dagger} = 1 \label{unitary}. 
\end{equation}
      Equations (\ref{Hecke-1}) to (\ref{CR}) show that the 
operator $e^{\dagger}$ satisfy the relations 
\begin{eqnarray}
    D_j e^{\dagger} - e^{\dagger} D_{j+1}
&=& 0, \quad (j=1,\cdots,N-1) \label{Permutation-1} \\
    D_N e^{\dagger} - e^{\dagger} D_1 \label{Permutation-2}
&=& e^{\dagger}.
\end{eqnarray}
      That is, $e^{\dagger}$ displaces all the subscripts of $D_i$ 
by one periodically. 
      Equations (\ref{Hecke-1}) to (\ref{CR}) also deduce the relation
among $e^{\dagger}$ and \{$X_{i,i+1}$\}: 
\begin{eqnarray}
      X_{i,i+1} e^{\dagger} &=& e^{\dagger} X_{i+1,i+2},
      \quad (i=1,\cdots,N-2) \label{Displace-1} \\
      {}X_{N-1,N} (e^{\dagger})^2 &=& (e^{\dagger})^2 X_{12}.
   \label{Displace-2}
\end{eqnarray}
      These equations show that $e^{\dagger}$ also displaces 
all the subscripts of the braid-exclusion operators by one. 


      Before constructing raising and lowering operators, 
we introduce an operator 
\begin{equation}
a_i^{\dagger}
= X_{i,i+1} X_{i+1,i+2} \cdots X_{N-1,N} e^{\dagger} 
\quad (i=1,\cdots,N)
\end{equation}
as an intermediate. 
      In the case of $i=N$ this equation reads as 
$a_N^{\dagger} = e^{\dagger}$. 
      We call $a_i^{\dagger}$ the {\it constituent operator}. 
      The constituent operators and the pseudomomenta satisfy 
the relations: 
\begin{eqnarray}
     D_j a_i^{\dagger} - a_i^{\dagger} D_{j+1}
&=& 0, \quad (1\leq j \leq i-1) \label{Creation-1} \\
    D_i a_i^{\dagger} - a_i^{\dagger} D_1
&=& a_i^{\dagger}, \label{Creation-2} \\
    {}[ D_j, a_i^{\dagger}]
&=& 0, \quad (i+1 \leq j \leq N) \label{Creation-3}
\end{eqnarray}
which are derived from (\ref{GEO-1}) to (\ref{GEO-3}), 
(\ref{Permutation-1}) and (\ref{Permutation-2}).
      The constituent operators and the braid-exclusion operators 
satisfy the relations: 
\begin{eqnarray}
X_{i,i+1} a_{i+1}^{\dagger} &=& a_i^{\dagger}, \label{Anyon-1} \\
X_{i,i+1} a_j^{\dagger} &=&
 \left\{
     \begin{array}{rl}
       a_j^{\dagger} X_{i+1,i+2}, & \quad (j \geq i+2) \\
       a_j^{\dagger} X_{i,i+1}, & \quad (j \leq i-1)
     \end{array}
 \right. \label{Anyon-2} \\
 a_i^{\dagger} a_j^{\dagger}
 &=& a_j^{\dagger} a_{i+1}^{\dagger} X_{12}, \quad (j \geq i+1)
\label{Anyon-3}
\end{eqnarray}
which are derived from (\ref{Braid-1}), (\ref{Braid-2}), 
(\ref{Displace-1}) and (\ref{Displace-2}). 
      Number-like operators $a_i^{\dagger} a_i$ and $a_i a_i^{\dagger}$ 
are expressed in terms of the pseudomomenta as follows: 
\begin{eqnarray}
& & a_i^{\dagger} a_i
 = \prod_{m=i+1}^N
     \left[
           (D_i-D_m)^2 - \beta^2
     \right], \quad (1 \leq i \leq N-1) \label{Number-1} \\
& & a_i a_i^{\dagger}
 = \prod_{m=i+1}^N
     \left[
           (D_1-D_m+1)^2 - \beta^2
     \right], \quad (1 \leq i \leq N-1) \label{Number-2} \\
& & a_N^{\dagger} a_N = a_N a_N^{\dagger} =1, \label{Number-3} 
\end{eqnarray}
which are derived from (\ref{GEO-1}) to (\ref{Exclusion}) and 
(\ref{unitary}) to (\ref{Permutation-2}).


      The {\it raising operator} is defined as a simple power of
 a constituent operator: 
\begin{equation}
       b_i^{\dagger} = (a_i^{\dagger})^i , \quad (i=1, \cdots, N) 
\end{equation}
and the corresponding {\it lowering operator} is its hermitian conjugate. 
      The raising operators and the pseudomomenta satisfies 
the commutation relations: 
\begin{equation}
       [D_i, b_j^{\dagger}] = \theta (j-i) b_j^{\dagger} , 
\label{commDb}
\end{equation}
as is derived from (\ref{Creation-1}) to (\ref{Creation-3}). 
      That is, $b_j^{\dagger}$ raises by one the eigenvalues of 
pseudomomenta with subscript $i$ for $i \le j$ 
and is qualified to be called a raising operator. 
      The raising operators are boson-like since they commute 
with each other: 
\begin{equation}
      [b_i^{\dagger}, b_j^{\dagger}] = 0,
\end{equation}
which are derived from (\ref{Anyon-1}) to (\ref{Anyon-3}). 

      Number-like operators are expressed in terms of the 
pseudomomenta as: 
\begin{eqnarray}
      & & b_i^{\dagger} b_i = \prod_{l=1}^i \prod_{m=i+1}^N
 \left[
      (D_l-D_m)^2 - \beta^2  
 \right], \quad (1 \leq i \leq N-1) \label{Number-4} \\
      & & b_i b_i^{\dagger} = \prod_{l=1}^i \prod_{m=i+1}^N
 \left[
      (D_l-D_m+1)^2 - \beta^2
 \right], \quad (1 \leq i \leq N-1) \label{Number-5} \\
      & & b_N^{\dagger} b_N = b_N b_N^{\dagger} =1 , 
\label{Number-6}
\end{eqnarray}
which are derived from (\ref{Creation-1}) to (\ref{Creation-3}) and  
(\ref{Number-1}) to (\ref{Number-3}). 
Further (\ref{Anyon-1}) to (\ref{Anyon-3}) yields the following 
relations: 
\begin{eqnarray}
      X_{i,i+1} b_j^{\dagger}
         &=& b_j^{\dagger} X_{i,i+1},\quad (i \neq j) \\
      b_i^{\dagger} X_{i,i+1} b_i^{\dagger}
         &=& [(D_{i+1}-D_i+1)^2-\beta^2] X_{i,i+1}
      b_{i-1}^{\dagger} b_{i+1}^{\dagger}.
\end{eqnarray}


      We now construct the Fock space of the Sutherland model 
by using the set of operators \{$D_i,b_j,X_{kl}$\}. 
      Concretely, we produce all the eigenstates of $\{D_i\}$, 
starting from a state and multiplying operators \{$b_j,X_{kl}$\} 
to it. 
      These states are also eigenstates of the Hamiltonian because of 
the commutability of $H$ and $\{D_i\}$. 
       An eigenstate with different energy level is produced 
by multiplying the raising or lowering operators, and 
a degenerate state is produced by multiplying the braid-exclusion 
operators. 

      We label an eigenstate of \{$D_i$\} by their eigenvalues 
\{$k_i$\} as
\begin{equation}
      D_i \mid k_1, k_2, \cdots, k_N > 
      = k_i \mid k_1, k_2, \cdots, k_N > .
      \quad (i=1, \cdots, N) 
\label{eigenstate}
\end{equation}
      We start the construction with a state which has the 
eigenvalues $k_i = \alpha_i$ $(i=1,\cdots,N-1)$ and is annihilated 
by lowering operators as 
\begin{equation}
      b_i \mid \alpha_1, \alpha_2, \cdots, \alpha_N > = 0 . 
      \quad (i=1,\cdots,N-1) 
\label{groundstate}
\end{equation}
      The case of $i=N$ is excluded in this equation, since $b_N (= e^N)$ 
is exceptionally unitary and does not annihilate any state. 
     Equation ($\ref{groundstate}$) reduces to 
\begin{equation}
      X_{i,i+1} \mid \alpha_1, \alpha_2, \cdots, \alpha_N > = 0 
      \quad (i=1,\cdots,N-1) 
\label{vacuum-2}
\end{equation}
due to the definitions of $a_i$ and $b_i$. 
      We begin the construction of the Fock space with a state 
satisfying condition $\alpha_1 > \alpha_2 > \cdots > \alpha_N$. 
     Then (\ref{vacuum-2}) reduces to 
\begin{equation}
      K_{i,i+1} \mid \alpha_1, \cdots, \alpha_N >
      = {\rm sgn}(\beta) \mid \alpha_1, \cdots, \alpha_N > ,
\end{equation}
by using the definition (\ref{definitionX}) of $X_{i,i+1}$ and 
the algebra, (\ref{Hecke-1}) to (\ref{Hecke-3}). 
      Hence the state $\mid \alpha_1, \cdots, \alpha_N >$ is a 
symmetric (antisymmetric) function for $\beta>0$ ($\beta<0$). 

      To examine possible values of \{$\alpha_i$\}, we operate 
another $X_{i,i+1}$ to (\ref{vacuum-2}). 
      Then we see that \{$\alpha_i$\} are related to each other 
since (\ref{vacuum-2}) and (\ref{Exclusion}) yields the relation 
$(\alpha_i - \alpha_{i+1})^2 = \beta^2$. 
      In reality there stands a stronger condition: 
\begin{equation}
      \alpha_i - \alpha_{i+1} = |\beta| , \quad (i=1,\cdots,N-1)
\end{equation}
which is obtained by a calculation with (\ref{Hecke-1}) to 
(\ref{Hecke-3}). 
      This condition is rewritten as
\begin{equation}
      \alpha_i = \alpha_0 + \frac{N+1-2i}{2} |\beta| 
      \quad (i=1,\cdots,N) 
\end{equation}
with undetermined constant $\alpha_0$ $(-1/2 < \alpha_0 < 1/2)$. 
      This kind of undetermined constant always appears 
in quantum mechanics on $S^1$ \cite{Ohnuki}. 
      Hereafter we choose it as $\alpha_0 = 0$ so that the total 
momentum $P$ of this state vanishes. 
      We write the state with $\alpha_0 = 0$ in (\ref{groundstate}) 
simply as $\mid 0 >$: 
\begin{equation}
      \mid 0 > \equiv \Big| \frac{N-1}{2}|\beta|, \frac{N-3}{2}|\beta|, 
      \cdots, -\frac{N-1}{2}|\beta| \Big> . 
\label{groundstate2}
\end{equation}
      We will see that this state is the true ground state in the Fock 
space which we are going to construct. 

      We have a series of excited states when we operate raising 
 operators to the ground state $\mid 0 >$. 
      By introducing a new notation, we write them as follows: 
\begin{equation}
      \mid n_1, n_2,\cdots,n_N \gg 
      \equiv (b_1^{\dagger})^{n_1-n_2}(b_2^{\dagger})^{n_2-n_3}
          \cdots (b_N^{\dagger})^{n_N} \mid 0 > . 
\label{excited_state}
\end{equation}
      Here we must impose the constraint $n_i \ge n_{i+1}$ 
$(i=1,\cdots, N-1)$ so that the power of $b_i^{\dagger}$ is positive; 
      $b_i^{\dagger}$ ($i \ne N$) generally has no inverse operator 
since $b_i^{\dagger} b_i$ has eigenvalue 0 as seen in 
(\ref{Number-3}). 
      In contrast the power $n_N$ of the last operator $b_N^{\dagger}$ 
is unrestricted because of its unitarity (\ref{Number-6}). 
      The negative power of $b_N^{\dagger}$ is read as 
the positive power of $b_N$: i.e. $(b_N^{\dagger})^{n} = (b_N)^{-n}$. 
      The constraint is concisely written as 
\begin{equation}
      n_1 \geq n_2 \geq \cdots \geq n_N . 
\label{condition_n}
\end{equation}
      The states defined by (\ref{excited_state}) are eigenstates of 
the pseudomomenta as is shown by (\ref{commDb}): 
\begin{equation}
      D_i \mid n_1, n_2,\cdots,n_N \gg
      = \left(
            n_i + \frac{N+1-2i}{2}|\beta|
      \right) \mid n_1, n_2,\cdots,n_N \gg . 
\end{equation}
      Hence $\mid n_1, n_2,\cdots,n_N \gg$ is identified as 
\begin{equation}
      \mid n_1, n_2,\cdots,n_N \gg = \mid k_1, k_2, \cdots, k_N >
      \label{definition}
\end{equation} 
with eigenvalue $k_i = n_i + (N+1-2i)|\beta|/2$ for $D_i$ 
($i=1,\cdots,N$). 
      The norm of this state is calculated as 
\begin{eqnarray}
& &   \ll n_1, \cdots ,n_N \mid n_1, \cdots , n_N \gg \nonumber \\
& & = \prod_{i=1}^{N-1} \prod_{l=1}^i
   \prod_{m=i+1}^N \prod_{r=1}^{n_i-n_{i+1}} 
\left[
  \left(
         (m-l)\beta + r + n_{i+1} - n_m
  \right)^2 - \beta^2  
\right]
\end{eqnarray}
by means of the relations (\ref{commDb}) to (\ref{Number-6}). 

      Next we operate a braid-exclusion operator $X_{i,i+1}$ to 
the eigenstate (\ref{eigenstate}) of \{$D_i$\}. 
     Then the relations (\ref{GEO-1}) to (\ref{Exclusion}) yields 
the following equation: 
\begin{equation}
      X_{i,i+1} \mid \cdots,k_i,k_{i+1},\cdots > 
      = \sqrt{(k_{i+1}-k_i)^2-\beta^2} 
           \mid \cdots,k_{i+1},k_i,\cdots >. 
\label{excited_X}
\end{equation}
       Hence, if $|k_{i+1} - k_i| \ne |\beta|$, 
$X_{i,i+1}$ produces a new state in which eigenvalues 
$k_i$ and $k_{i+1}$ are exchanged. 
      The equation (\ref{excited_X}) for states corresponds to 
the relation (\ref{GEO-1}) and (\ref{GEO-2}) for operators, 
which means the exchange of $D_i$ and $D_{i+1}$. 
      Operating \{$X_{i,i+1}$\} to $\mid k_1, k_2, \cdots, k_N >$ 
in (\ref{definition}) finite times, we reach any possible order of
 \{$k_i$\}. Redefining $k_i$ as the eigenvalue of $D_i$, 
possible eigenvalues of \{$D_i$\} are written as
\begin{equation}
      k_i = n_{\sigma(i)} + \frac{N+1-2\sigma(i)}{2}|\beta| , 
      \quad (i=1,\cdots,N) 
\end{equation}
where $\sigma$ is a permutation among 1 to {\it N} 
which satisfies $n_{\sigma(i)} \ne n_{\sigma(j)}$
 for $| \sigma(i) - \sigma(j) | = 1 $.
      For $|\beta| = 1$, the constraint (\ref{condition_n}) 
is equivalent to the Pauli principle: $| k_i - k_j | \ne 1$. 
      Hence for any $\beta$ the constraint (\ref{condition_n}) 
describes a generalized Pauli principle: 
\begin{equation}
      \mid k_i - k_j \mid \geq |\beta|. 
\label{Pauli-3}
\end{equation}

      When a set of pseudomomentum eigenvalues \{$k_i$\} is known, 
(\ref{Hamiltonian-3}) gives the energy eigenvalue as 
\begin{equation}
      E = \sum_{i=1}^{N} k_i^2 . 
\end{equation}
      This equation shows that the set of $k_i = \alpha_i$ 
($i = 1, 2, \cdots , N$) gives the lowest energy and 
(\ref{groundstate2}) is the true ground state. 
      In an arbitrary set \{$k_i$\}, the energy $E$ is invariant under 
an exchange of $k_i$'s. 
      The exchanged set gives a state with the same energy as 
the original if $|k_i - k_j| \ne |\beta|$ 
($i \ne j$). 
      Thus the braid-exclusion operators \{$X_{i,i+1}$\} create 
degenerate states by repeating (\ref{excited_X}). 
      The ground state is not degenerate, since the operation of 
\{$X_{i,i+1}$\} to the ground state ($k_i = \alpha_i$) gives 0 
due to (\ref{excited_X}). 

      The degeneracy of an energy eigenvalue is given by counting the 
number of possible combinations of the corresponding set \{$k_i$\}. 
      We take out all the quantum numbers $m_1$, $m_2$, $\cdots$, 
$m_L$ which are included in \{$n_j$\} and are different from 
each other. 
      Then we define $l_i$ for each $m_i$ so that $l_i$ is the number 
of elements equal to $m_i$ in \{$n_j$\}. 
      In terms of \{$l_j$\} the degeneracy is given by 
\begin{equation}
      \frac{N!}{l_1!l_2!\cdots l_L!} . 
\end{equation}
      Thus we have reproduced all the eigenenergies and their 
degeneracy for the Sutherland model.


      In summary, we have found a novel algebraic formalism for 
the eigenvalue problem of the Sutherland model. 
      All the energy eigenstates are obtained as eigenstates of 
pseudomomenta \{$D_i$\}. 
      The formalism is based on raising operators \{$b_i^\dagger$\} 
and braid-exclusion operators \{$X_{i,i+1}$\} as well as 
pseudomomenta \{$D_i$\}. 
      While $b_i^\dagger$ creates another state with different energy, 
$X_{i,i+1}$ creates another degenerate state. 
      The calculation of the correlation function in the present 
formalism is a future problem. 

      We would like to thank Yoshio Ohnuki and Shinsaku Kitakado 
for useful discussions.


\end{document}